# Evaluation of imaging protocol for ECT based on CS image reconstruction algorithm

ZHOU Xiao-Lin(周小林)[1,2,3] YUN Min-Kai(贠明凯)[1,2] CAO Xue-Xiang(曹学香)[1,2] LIU Shuang-Quan(刘双全)[1,2]

WANG Lu(王璐)[1,2] HUANG Xian-Chao(黄先超)[1,2] WEI Long(魏龙)[1,2;1)]

[1] Key Laboratory of Nuclear Analytical Techniques, Institute of High Energy Physics, Chinese Academy of Sciences, Beijing 100049, China

[2] Beijing Engineering Research Center of Radiographic Techniques and Equipment, Beijing 100049, China

[3] Graduate University of Chinese Academy of Sciences, Beijing 100049, China

**Abstract**:

SPECT (Single-photon Emission Computerized Tomography) and PET (Positron Emission Tomography) are essential medical imaging tools, for which the sampling angle number, scan time should be chosen carefully to compromise between image quality and the radiopharmaceutical dose. In this study, the image quality of different acquisition protocol was evaluated via varied angle number and count number per angle with Monte Carlo simulation data. It was shown that when similar imaging counts were used, the factor of acquisition counts was more important than that of the sampling number in ECT (Emission Computerized Tomography). To further reduce the activity requirement and the scan duration, an iterative image reconstruction algorithm for limited-view and low-dose tomography based on compressed sensing theory has been developed. The total variation regulation was added in the reconstruction process to improve SNR (Signal to Noise Ratio) and reduce the artifacts caused by the limited angle sampling. Maximization of maximum likelihood of the estimated image and the measured data and minimization of the total variation of the image are alternative implemented. By using this advanced algorithm, the reconstruction process is able to achieve image quality matching or exceeding that of normal scan with only half of the injection radiopharmaceutical dose.

## 1. INTRODUCTION

SPECT and PET are established diagnostic tools widely appreciated in the clinical fields of oncology, neurology, cardiology and several others[1]. However, the lack of standardized acquisition protocols has been identified as a problem that limits their potential in the lesion detectability and diagnosis confidence. Feasibility study of optimizing acquisition protocol between angle sampling and activity requirement has been implemented along with the extension of the use of these imaging tools[2-4]. In general, the trade-off between image acquisition time and noise levels has determined the standard protocols and scan times. Images of better quality are obtained through larger number of sampling and more acquisition counts, which means more radionuclide or longer acquisition time. But in the clinical settings shorter acquisition time and less radionuclide dose are preferable if image quality is the same as that of normal scan. Shorter acquisition time is beneficial for patient's tolerance and allowing higher throughput for screening applications[5]. Less radionuclide dose reduce the risk of radiation exposure for patients and nuclear staff[6].

Several studies have suggested the importance of optimizing acquisition times or the injected doses of radiopharmaceuticals to improve the quality of images in ECT[7-9]. However, these studies offered recommendations on acquisition protocol only by traditional reconstruction algorithms such as FBP (Filtered Back Projection) and MLEM (Maximum Likelihood Expectation Maximization). These algorithms deliver unsatisfactory and noisy results in the shorter acquisition time and less radionuclide dose cases. To tackle this challenge, the CS (Compressed Sensing)

Supported by National Natural Science Foundation of China (81101175)

1) Email: weil@ihep.ac.cn

theory[10] is introduced in the image reconstruction. Moreover, several works have used the priori information of total variation to improve the quality of images in ECT reconstructions[11-13]. In this study, The CS based EM-TV (Expectation Maximization- Total Variation) algorithms is used to optimize the acquisition protocol. In the first part, the concentration is on the effect that how sampling angle and counts per angle impact on the image quality with both MLEM and EM-TV algorithms. Furthermore, in the second part, the EM-TV algorithm is particularly used to halve the acquisition time and the dose requirement while preserving the image quality.

## 2. METHODS AND MATERIALS

Since the reconstruction algorithms have considerable effects on the acquisition protocol optimization, both traditional and CS-based reconstruction methods were evaluated, namely the well-known MLEM and EM-TV, respectively. Our study had 2 parts. In the first, six sets of ECT simulate data with various angle numbers and counts per angle (with a fixed total imaging counts) were acquired to evaluate how these two factors affect the image quality. The two reconstruction algorithms were used and the SNR and CNR (Contrast-to-Noise Ratio) were calculated to evaluate the lesion detectability and diagnosis confidence. In the second part of the study, EM-TV algorithm was prospectively used to develop the image protocol of half-acquisition. In other words, with EM-TV algorithm, we expected to get the same or better image quality using shorter acquisition time and less view angels than that of MLEM with full-acquisition.

### 2.1 Reconstruction Algorithms

MLEM is a widely used iterative algorithm to maximize the expectation maximization likelihood function[14]. The significant merit of this algorithm is that it can achieve much better image quality compared with that of FBP [15].

The formula of MLEM algorithm is

$$f^{(k)}(i) = \frac{f^{(k-1)}(i)}{\sum p(i,j)} \sum \frac{p(i,j)d(j)}{\sum p(i',j)f^{(k-1)}(i')}, \quad (1)$$

where $f^{(k)}(i)$ is the $i$th element of the reconstructed image at the $k$th iteration, $p(i,j)$ is the system matrix which represents the probability of an event in pixel $i$ being detected by LOR (Line Of Response) $j$, $d(j)$ is the $j$th projection. System matrix is a key factor in MLEM algorithm which models the relationship between the reconstructed image and the projection data.

Common reconstruction algorithms including MLEM yield undesirable artifacts in the reconstructed images with limited view and low dose data. CS algorithm[10] is a well-established approach for signal recovery, which mainly relies on the sparsity recovery of the target signal. Based on the assumption that the target signal has a sparsifying form, CS algorithm has been acknowledged to show convincing expertise in dealing with the limited view and low dose cases[16, 17]. The L1-minimized method usually used to solve the constrained optimization problem in CS. The formula of L1-minimized method is

$$\min\|\Psi\vec{f}\|_1 \quad s.t. \quad M\vec{f} = \vec{g}, \quad (2)$$

where $\Psi$ is the sparsifying transform, $\vec{f}$ is the true image, M is the system matrix and $\vec{g}$ is the measured data. The gradient transform is widely used as a sparsifying transform in sparse-view image processing. Since medical images have the sparsity in the gradient transform[18], it will be possible to reconstruct the accurate image by recovering the sparsity. As the L1-norm of the gradient transform is the total variation, the expression of the L1-minimized method becomes

$$\min\|\vec{f}\|_{TV} \quad s.t. \quad M\vec{f} = \vec{g}. \quad (3)$$

A two-step iterative method[19] is used to solve Eq.(3). The first step is to enforce measured data to the true activity where the traditional reconstruction methods can be applied. The next step is to minimize the TV of the image. This total variation based algorithm has been recently investigated in Cone-Beam Computed Tomography (CBCT) as TV-POCS (Projection Onto Convex Sets) algorithm[20], which has been used to preserve edges, with the assumption that most images are piece-wise constant. However, the Poisson noise due to photon counting statistics in nuclear imaging may seriously disturb the TV-minimization. In this paper, the EM algorithm which is considered having superiority under a Poisson noise was used in the first step, and the gradient descent method was used in the second step. The execution step of this EM-TV algorithm is
the EM-step:

$$f_{EM}^k = \frac{f^{(k-1)}(i)}{\sum p(i,j)} \sum \frac{p(i,j)d(j)}{\sum p(i',j)f^{(k-1)}(i')}, \quad (4)$$

the TV-step:

$$f_{EMTV}^{k,l} = f_{EMTV}^{k,l-1} - a\frac{\vec{v}_{TV}^{k,l-1}}{\left|\vec{v}_{TV}^{k,l-1}\right|}, \quad (5)$$

$$where \quad \vec{v}_{TV}^{k,l-1} = \left.\frac{\partial\|\vec{f}\|_{TV}}{\partial f}\right|_{f=f_{EMTV}^{k,l-1}}, \quad (6)$$

and where $k$ is the iteration number of the EM-TV method, $l$ is the iteration number of the TV-step, $a$ is a relaxation factor to balance the two step. $f_{EMTV}^{k,0}$ in the TV-step should be set to $f_{EM}^k$ in the EM-step and $f_{EM}^{k-1}$ should be set to the output image of the TV-step in the last iteration. Since $a$ could be related to the view numbers, we set different value to $a$ in each experiment.

## 2.2 Simulation Model

A single head SPECT with a low-energy, high-resolution collimator was simulated as a typical ECT equipment by GATE (Geant4 Application for Tomographic Emission)[21]. The detector component of this equipment is composed of a $62\times62$ array of $2\times2\times6$ mm³ NaI crystals. According to the energy resolution of the system, the energy window was set at a 20% symmetric window at 140 Kev. To flexibly change the sampling angle, all of the scans were acquired with a circular orbit with step-and-shoot acquisition over 360°. Six experiments were set with fixed total counts in the first part, as shown in Table 1. In the second part, half-acquisition data was selected for some view numbers, compared with the full-acquisition for the same angle numbers, as shown in Table 2. The scatter events had been rejected before reconstruction.

Table 1 the six sets of experiments for full-acquisition of part 1

| Num. of sets | 1 | 2 | 3 | 4 | 5 | 6 |
|---|---|---|---|---|---|---|
| Angle number | 120 | 60 | 40 | 30 | 24 | 20 |
| Counts/angle ($\times 10^3$) | 10 | 20 | 30 | 40 | 50 | 60 |
| Reconstruction algorithm | Both MLEM and EM-TV | | | | | |

Table 2 the six sets of experiments for comparison of half- acquisition and full-acquisition of part 2

| | Half- acquisition | | | Full- acquisition | | |
|---|---|---|---|---|---|---|
| Num. of sets | 1 | 2 | 3 | 4 | 5 | 6 |
| Angle number | 60 | 30 | 20 | 60 | 30 | 20 |
| Counts/angle ($\times 10^3$) | 10 | 20 | 30 | 20 | 40 | 60 |
| Reconstruction algorithm | EM-TV | | | MLEM | | |

A cylinder phantom with internal diameter of 90mm was simulated in this study. The activity concentration of $^{99m}Tc$ in the phantom background was 0.2 μCi/CC, as a typical activity concentration in clinic. Inner the phantom, six small cylinder with the diameter of 18.5, 14, 11, 8.5, 6.5, 5mm were inserted and placed at a radial distance of 28.6mm from the center of the phantom. The two largest cylinders (18.5mm and 14mm) were filled with water containing no radioactivity for cold lesion imaging, whereas the four smallest cylinders (11mm, 8.5mm, 6.5mm, and 5mm) were filled with an activity concentration of 9:1 with respect to the background for hot lesion imaging (Fig.1).

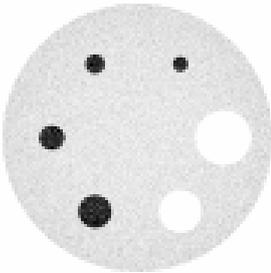

Fig.1 The transversal image of the phantom.

## 2.3 Evaluation Methods

A transverse image centered on the phantom was used for analysis and six circular ROIs (Regions of interest) were drawn over the center of the six inserted cylinders. Similarly, a circular ROI with the diameter of 30mm was drawn on the center of the image as a background ROI.

To evaluate the detection rate of lesions and the evaluation accuracy, the SNR and CNR were used as quality measurement.

$$\text{SNR} = \frac{M_{bg}}{STD_{bg}}, \qquad (7)$$

$$\text{CNR} = \frac{M_{hot}-M_{bg}}{STD_{bg}}, \qquad (8)$$

$$\text{CNR} = \frac{M_{bg}-M_{cold}}{STD_{bg}}, \qquad (9)$$

where $M_{hot}$ is the average of the hot ROI, $M_{clod}$ is the average of the cold ROI, $M_{bg}$ is the average of the background ROI, and $STD_{bg}$ is the standard deviation of the background ROI. The comparison of SNR and CNR were plotted for each algorithm and each experiment under different iteration times.

# 3 RESULTS

## 3.1 Full-acquisition Results of Part1

The six sets of simulated data (table 1) were reconstructed by MLEM algorithm and EM-TV algorithm with different iteration times. Although the sampling angles are fewer, it is apparent from Fig.2 that the uniformity has been visibly increased for both MLEM and EM-TV results with more counts per angle.

The SNR and CNR curves for measurement of the lesion detectability are shown in Fig.3. From the MLEM results, it can be seen clearly that a smaller view number could get higher SNR and CNR with a fixed total counts after a certain iteration times. The regularity was about the same for the EM-TV results. Moreover, it is to be observed that: (1) the SNR and CNR of the EM-TV results were higher than that of MLEM results, (2) the SNR and CNR of the 30-view-angle experiment by EM-TV was a little distinctive which caused by the effect of Poisson noise to the TV-minimization step, but there were signs that with the increase of iteration number, it will follow the regularity mentioned before finally.

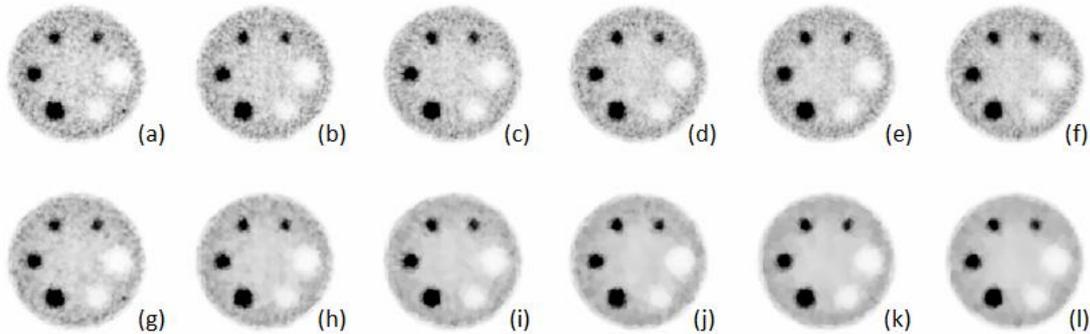

Fig.2 The transverse images in the center slice after 30 iterations of full-acquisition using MLEM (the top row) and EM-TV (the bottom row) with different view numbers and counts per angle in table 1. The total counts are fixed and the (view numbers, counts/angle) of each experiments is (a, g) (120, $10 \times 10^3$); (b, h) (60, $20 \times 10^3$); (c, i) (40, $30 \times 10^3$); (d, j) (30, $40 \times 10^3$); (e, k) (24, $50 \times 10^3$); (f, l) (20, $60 \times 10^3$).

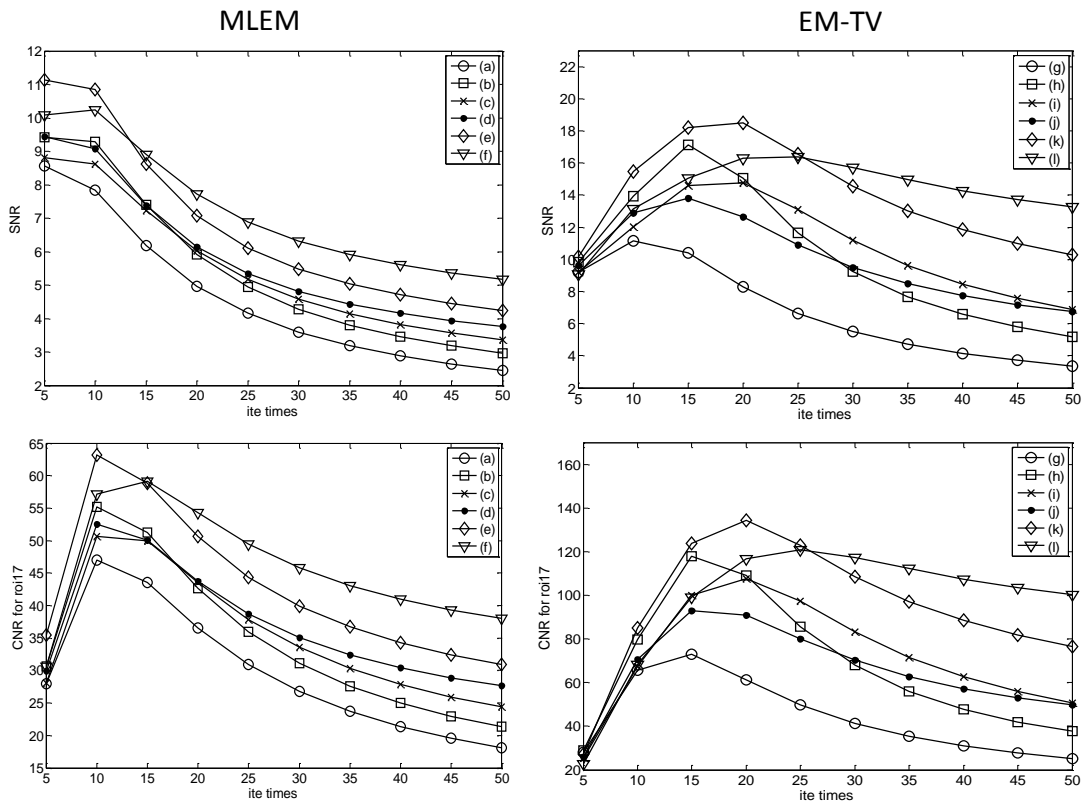

Fig.3 The SNR and CNR of ROI17 vs. iteration times for the images in Fig.2. The top row shows the SNR value; the bottom row shows the CNR value; the left rows shows the MLEM results and the right rows shows the EM-TV results.

## 3.2 Half- acquisition Results of Part2

To reduce the radiological dose and acquisition time, the images of comparative experiments in Table 2 are shown in Fig.4, and the SNR and CNR vs. iteration times are shown in Fig.5. Compared with (a) and (d), the SNR and CNR was about the same. Moreover, the SNR and CNR of (b) was higher than that of (e), and of (c) was much higher than that of (f). That is, with half counts/angle, EM-TV could gain greater advances in the case of less view numbers.

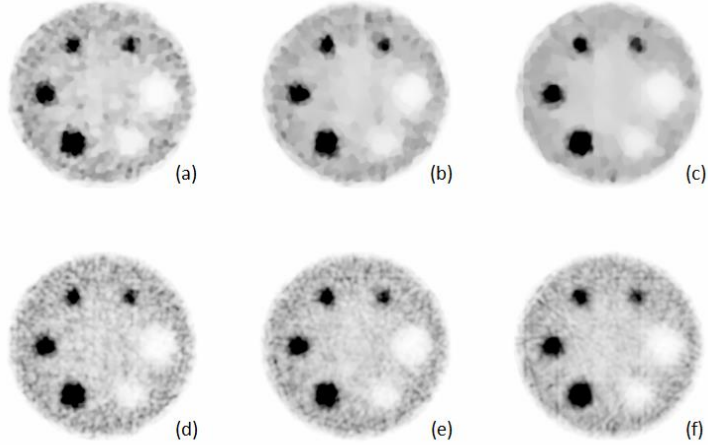

Fig.4 The transverse images in the center slice after 30 iterations with half-acquisition using EM-TV (the top row) and full-acquisition using MLEM (the bottom row) in Table 2. The (view numbers, counts/angle) and reconstruction algorithm of each experiments is: (a) $(60, 10 \times 10^3)$-EMTV, (b) $(30, 20 \times 10^3)$-EMTV, (c) $(20, 30 \times 10^3)$-EMTV, (d) $(60, 20 \times 10^3)$-MLEM, (e) $(30, 40 \times 10^3)$-MLEM, (f) $(20, 60 \times 10^3)$-MLEM.

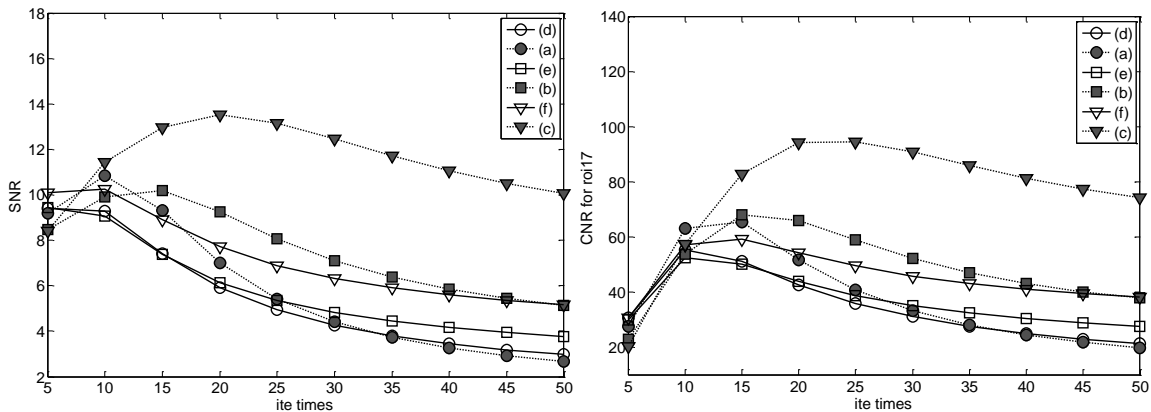

Fig.5 The CNR of roi17 vs. iteration times for the images in Fig.4. The left figure shows the SNR value; the right figure shows the CNR value.

## 4 DISCUSSION

In nuclear imaging, the radiation dose and imaging efficiency are mostly concerned, therefore in the first part of this study, the sampling angle and the counts per angle were compromised so that better image quality can be obtained. The results of the experiments presented in Table 1 using both MLEM and EM-TV algorithms (Fig.2, 3) revealed that the noise lever per angle plays more important in the image quality. This will provide useful information for optimization of the measurement chain.

To reduce scan time and tracer requirement, noisier images are allowed or more $\gamma$ detectors are introduced which result in inaccurate diagnosis and highly cost. The recently introduced iterative reconstruction algorithms incorporate noise regularization and resolution recovery may provide a new alternative. In the second part of this study, a novel compressed sensing-based reconstruction from significantly fewer measurements than traditionally required was presented, thus demonstrating potential of reduction in scan time and radiopharmaceutical doze with benefits for patients and health care economics. Several metrics, such as SNR and CNR, are used to compare the performance of the developed method and traditional reconstruction algorithm. It is indicated by Fig.4 and Fig.5 that the new approach allows the same quality images for the view number of 60, and higher quality images for the view number of 30 and 20. In other words, the image quality can be preserved or improved, even if the radiopharmaceutical injection dose and the scan time are reduced, which does not only help to reduce the harmful radiation dose exposed to the patients and the staff, but also enhance the scanner efficiency. The reduction of acquisition time would also lead to fewer motion artifacts according with greater scanner efficiency.

Another way to improve the performance of imaging system is to incorporate the characteristic of the detector response in the reconstruction process as a resolution recovery algorithm. This work is in progress.

## 5 CONCLUSION

Under similar imaging counts, acquisition counts per angle should be considered more important than the sampling number in ECT. In addition, by using the CS-base EM-TV algorithm, the injected dose to the patient can be halved while obtaining even better or at least the same image quality compared with a full dose scan.

# 基于压缩感知重建算法的ECT采集方案评估

**摘要：**

单光子发射断层扫描仪（SPECT）和正电子发射断层扫描仪（PET）是医学中常用的成像设备，在成像过程中，为了平衡图像质量和放射性剂量，需要对采集角度数和采集时间进行权衡。在本项研究中，我们基于蒙特卡洛模拟数据，对不同采样条件，主要是不同角度数和每角度采集时间情况下的图像质量做了评估。结果表明，对于发射单光子计算机断层扫描仪（ECT）来说，在总采集计数一定的情况下，每角度采集计数比角度数发挥更重要的作用。为了进一步降低剂量，减少采集时间，采用了一种基于压缩感知（CS）的迭代重建算法，并在其中加入全变分约束，用来提高图像信噪比以及减少由少角度采样带来的图像伪影。在此种重建算法中，最大化估计图像的似然函数以及最小化图像的全变分被迭代执行，效果是在总剂量减半的情况下，可以达到同样或者更好的图像质量。

**关键词：** 发射单光子计算机断层扫描仪；采集方案；压缩感知；全变分